\documentclass[final]{jfm}

\usepackage{graphicx}
\usepackage{newtxtext}
\usepackage{newtxmath}
\usepackage{natbib}
\usepackage{hyperref}
\usepackage{appendix}
\hypersetup{
    colorlinks = true,
    urlcolor   = blue,
    citecolor  = black,
}

\newcommand{\RomanNumeralCaps}[1]
\linenumbers

\title{Dynamics of rotating helices in viscous fluid}

\author{Chijing Zang\aff{1},
  Luke Omodt\aff{2},
  Moumita Dasgupta\aff{2},
 \and Xiang Cheng\aff{1,3}\corresp{\email{xcheng@umn.edu}}}

\affiliation{\aff{1}Department of Chemical Engineering and Materials Science, University of Minnesota, Minneapolis,
MN 55455, USA
\aff{2}Department of Physics, Augsburg University, Minneapolis,
MN 55455, USA
\aff{3}Saint Anthony Falls Laboratory, University of Minnesota, Minneapolis, MN, 55414, USA
}

\begin{document}
\maketitle

\begin{abstract}
We investigate the dynamics of a pair of rigid rotating helices in a viscous fluid, as a model for bacterial flagellar bundle and a prototype of microfluidic pumps. Combining experiments with hydrodynamic modeling, we examine how spacing and phase difference between the two helices affect their torque, flow field and fluid transport capacity at low Reynolds numbers. Hydrodynamic coupling reduces the torque when the helices rotate in phase at constant angular speed, but increases the torque when they rotate out of phase. We identify a critical phase difference, at which the hydrodynamic coupling vanishes despite the close spacing between the helices. A simple model, based on the flow characteristics and positioning of a single helix, is constructed, which quantitatively predicts the torque of the helical pair in both unbounded and confined systems. Lastly, we show the influence of spacing and phase difference on the axial flux and the pump efficiency of the helices. Our findings shed light on the function of bacterial flagella and provide design principles for efficient low-Reynolds-number pumps.
\end{abstract}


\section{Introduction}
\label{sec:headings}
Peritrichous bacteria, such as \textit{Escherichia coli} (\textit{E. coli}), swim by combining multiple independently driven flagella into a single rotating helical bundle, which generates hydrodynamic thrust and propels the cells forward \citep{Berg2004}. While the propulsion mechanism enabled by a single helical bundle is well studied \citep{Lauga_2020_book}, fluid-mediated interactions among multiple flagella within the bundle are still far from fully understood \citep{Bianchi_2023_flagella}. Although theoretical and numerical modeling have been empolyed to elucidate the collective dynamics of multiple flagella during bundle formation and within a functional bundle \citep{Kim2004_1, reichert2005, flores2005, janssen2011, reigh2012, kanehl2014, man2017, Chamolly2020, Tatulea2021, Tatulea2022, Park2024}, direct experimental validation of predicted flagellar dynamics remains difficult due to the small spatial scale ($\sim 20$ nm) and rapid time scale ($< 10$ ms) associated with the flagellar process \citep{Turner_2000_flagella}. To tackle this challenge, scaled experimental models have been developed to mimic flagellar dynamics at macroscopic scales \citep{Macnab_1977_flagella,Kim2003,Kim2004,Danis2019, Lim_2023_robot}. Most of these works, however, focused on the formation of the flagellar bundle. Few experiments have been conducted exploring the hydrodynamic interactions between multiple flagella and examining their consequence on the dynamics of flagella in a stable bundle. 

Beyond their function in bacterial motility, rotating helices can be harnessed as pumps and mixers in microfluidics. Immobilized bacteria in bacterial carpets generate strong near-surface flows, enabling fluid transport and mixing \citep{darnton2004,kim2008} and guiding particle motion along designed trajectories \citep{gao2015}. The possibility of using rotating helices as microfluidic pumps has also been explored in simulations \citep{martindale2017,dauparas2018,buchmann2018,rostami2022}. Artificial helical microrobots actuated by external fields have also been exploited for fluid pumping and mixing \citep{zhang2010, tottori2012}. These studies demonstrate the great potential of using collectively rotating helices for precise control of low-Reynolds-number flow in microfluidics. A deeper understanding of the hydrodynamic interactions between rotating helices and a detailed mapping of the surrounding flow characteristics become essential to fully realizing this potential.

Our work, integrating experiments and hydrodynamic modeling, addresses this critical knowledge gap and technical challenge and provides an in-depth study of the dynamics of a pair of rigid rotating helices in low-Reynolds-number ($\Rey$) flows.

\section{Methods}

\subsection{Experimental Setup}
\begin{figure}
  \centerline{\includegraphics[scale=0.42]{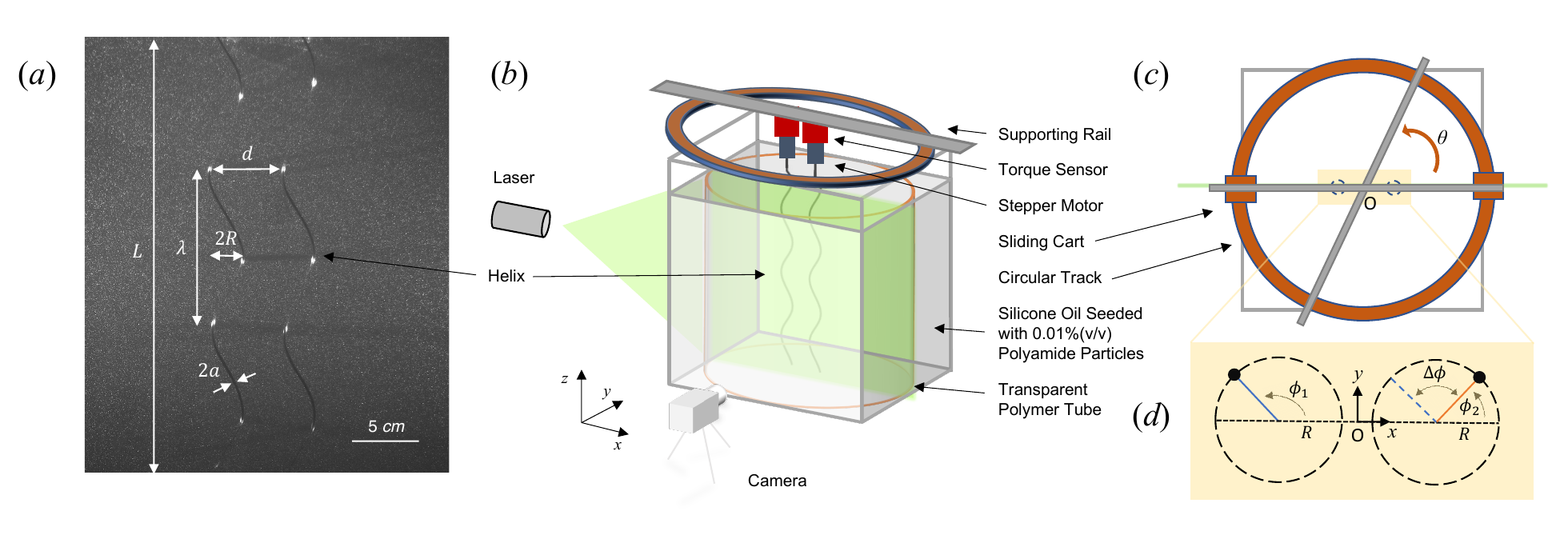}}
  \captionsetup{width=1\linewidth,justification=justified}
  \caption{Experimental setup. (a) Image of the two helices in silicon oil seeded with PIV tracers. The helical radius $R$, axial length $L$, pitch length $\lambda$, filament radius $a$, and the inter-flagellar spacing $d$ are marked. (b, c) Side and top views of the experimental apparatus. Stepper motors are mounted on a supporting rail attached to two sliding carts, which can rotate along a circular track by an angle $\theta$. (d) Schematic showing a plane, normal to the axes of two helices with the phase difference $\Delta \phi$. Dashed circles indicate the trajectories of the helices in the plane.}
\label{fig:setup}
\end{figure}

We construct a macroscopic model of bacterial flagella. The scaled helical flagella are left-handed and fabricated by wrapping stainless steel wires (radius $a = 1$ mm) around a cylindrical rod with a radius of $R = 1$ cm at a pitch length $\lambda = 12$ cm (Fig.~\ref{fig:setup}(a)). The helical axial length submerged in fluid is fixed at $L$ = 32 cm. The ratios $R/L = 0.026$ and $\lambda/L = 0.32$, closely matching those of \textit{E. coli} flagella at $R/L = 0.028$ and $\lambda/L = 0.35$ \citep{Turner_2000_flagella}. 

The schematic of our experimental setup is shown in Figs.~\ref{fig:setup}(b-d). The helices are inserted vertically into silicone oil (Clearco, density $\rho = 0.970$ g/cm$^3$, viscosity $\eta = 105$ Pa$\cdot$s), contained in a plastic cubic tank of a side length of $50R$. To impose a cylindrical boundary, we submerge a thin PET tube of radius $R_0 = 25R$ with a refractive index close to that of silicone oil in the tank. The axes of the two helices define an axial plane (the $x$-$z$ plane). The origin `O' of the orthogonal $x$-$y$ plane is set at the middle point connecting the axes of the two helices (Fig.~\ref{fig:setup}(c)). The axis of the tube is carefully aligned within the axial plane through the origin. Two separate stepper motors (NEMA 8) drive the rotation of the helices at a constant angular speed $\omega$ = 0.5 rad/s in the clockwise direction when looking from above, which sets a velocity scale $\omega R$ = 0.5 cm/s with $\Rey = \rho\omega R^2/\eta \approx 4 \times 10^{-4}$. The distance between the axes of the two helices $d$ and their phase difference $\Delta\phi$ are varied in our experiments. Two reaction torque sensors (TFF 400, FUTEK), one for each helix, are mounted on a supporting rail and connected to the stepper motors, which measure the torque applied by the motors to the helices. The supporting rail can rotate on a circular track concentric with the tube axis, enabling us to adjust the orientation of the axial plane $\theta$.

\subsection{PIV analysis and SBT model}
We use particle image velocimetry (PIV) to measure the 3D flow field of the rotating helices (PIVLab, Matlab). Silicone oil is seeded with polyamide microparticles (LaVision, $\rho = 1.03$ g/cm$^3$, diameter 60 $\mu$m) at a concentration of 0.01$\%$ (v/v). Two 50 mW, 532 nm continuous line lasers (CivilLaser), positioned on opposite sides of the tank, generate an aligned laser sheet approximately 1-mm thick in the measurement window through the tube axis. By varying the orientation of the axial plane with respect to the laser sheet, we image the axial flow at different polar angles $\theta$ (Fig.~\ref{fig:setup}(c) and Movie 1). The flow in the $x$-$y$ plane is captured from the bottom of the tank using a horizontal laser sheet and a 45-degree inclined mirror (Movie 2). Videos are recorded at 5 fps using a CMOS camera (Basler acA2040-90um). The spatial resolution of PIV is 1.3 mm in the $x$-$y$ plane and 4.4 mm in the $x$-$z$ plane. The root mean square velocity error is $0.03 \omega R$. We also simulate the flow field of two rotating helices using slender-body theory (SBT). The SBT formulas from \citet{rodenborn2013} for a single helix are extended for two rotating helices with different $d$ and $\Delta \phi$ (Appendix~\ref{appendix:SBT}).

\section{Results}

\subsection{Torque}
\begin{figure}
    \centerline{\includegraphics[scale=0.3]{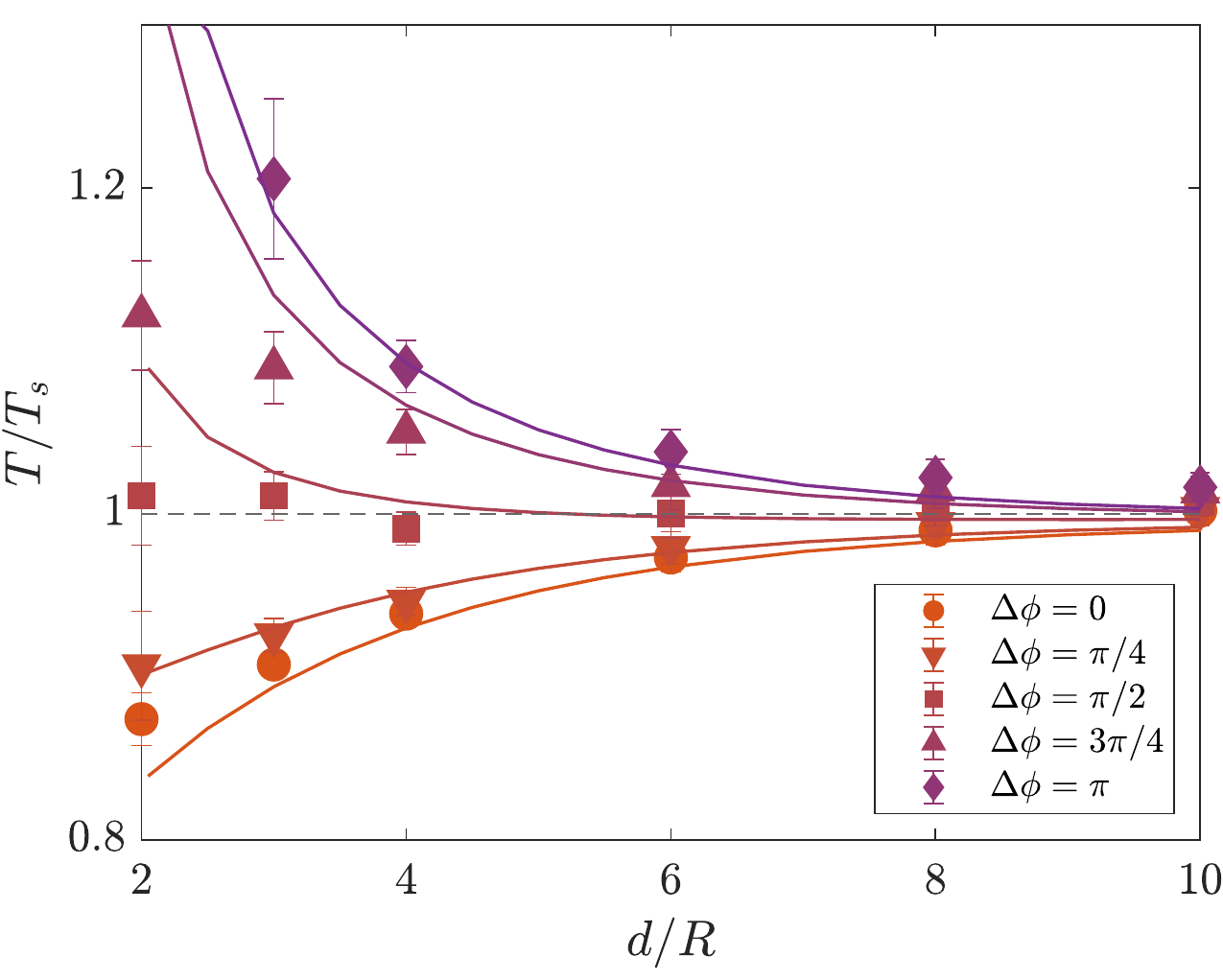}}
    \captionsetup{width=1\linewidth,justification=justified}
    \caption{The torque required to rotate one helix in a helical pair, $T$, as a function of the spacing between the helices, $d$, at different phase differences, $\Delta \phi$. $T$ is normalized by the torque required to rotate a single helix positioned at the center of the system $T_s$, and $d$ is normalized by the helical radius, $R$. Symbols are experimental data, while the lines show the corresponding SBT calculations.}
\label{fig:torque}
\end{figure}

We first measure the time-averaged torque required to rotate a single helix in a helical pair, where two helices, separated by distance $d$, rotate at the same speed with a phase difference $\Delta \phi$ (Fig.~\ref{fig:torque}). The torque is normalized by $T_s$, the torque required to rotate a single helix at the center of the tube at the same speed. At large separations, $T/T_s \to 1$, indicating negligible hydrodynamic interactions between distant helices. As the helices move closer, $T$ displays stronger deviation from $T_s$: $T < T_s$ for $\Delta \phi < \pi/2$, whereas $T  > T_s$ for $\Delta \phi > \pi/2$. Interestingly, the hydrodynamic coupling between the helices vanishes at a critical phase difference $\Delta \phi = \Delta \phi_c \approx \pi/2$, leaving $T = T_s$ regardless of their spacing. The SBT calculation in an unbounded fluid quantitatively matches our experiments (Fig.~\ref{fig:torque}), except at small $d$, where higher-order near-field interactions ignored in SBT become important due to the close proximity of the helices. The overall agreement suggests that the system boundary at $R_0 = 25R$ has a minimal effect. 

While the impact of hydrodynamic interactions on the torque of a flagellar bundle has been explored in simulations \citep{Kim2004_1, buchmann2018,Tatulea2021}, we provide direct experimental evidence of this nontrivial effect, which has significant implications for the dynamics of bacterial flagellar bundles. In particular, during bundle formation, flagella must synchronize their rotation ($\Delta \phi \to 0$) while reducing their separation ($d \to 0$). As the torque and the rotation speed of a helix are linearly correlated in Stokes flow via a drag coefficient $\xi$, $T=\xi\omega$, our results show that $\xi$ of a rotating helix in a helical pair decreases with decreasing $d$ and $\Delta \phi$, leading to the observed decrease in torque when the helices rotate at a constant speed (Fig.~\ref{fig:torque}). For bacterial flagella that operate under constant torque \citep{Berg2004}, this decrease in drag coefficient results in an increase in flagellar rotation speed $\omega=T/\xi$ as $d$ and $\Delta \phi \to 0$ during bundle formation. Notably, $\xi$ depends only on the geometry of the system and the fluid viscosity, regardless of whether the helices rotate at a constant speed or constant torque. Hence, while the power required to rotate a helix, $T\omega = \xi\omega^2$, decreases with $\xi$ as $d$ and $\Delta \phi$ reduce in our case with constant $\omega$, $T\omega = T^2/\xi$ increases with decreasing $\xi$ for bacteria during bundle formation under constant $T$. Thus, unlike in-phase beating of two undulating sheets, which minimizes energy dissipation \citep{Taylor1951}, flagellar synchronization is not a power-saving mechanism for bacteria \citep{reichert2005,kanehl2014}. Our results also indicate that increasing the number of flagella in a bundle reduces the drag coefficient of each flagellum in the bundle, allowing flagella to rotate at higher speeds under constant torque \citep{Kamdar_2023_multipleflagella}. 

\begin{figure}
\centering
\includegraphics[width=1\linewidth]{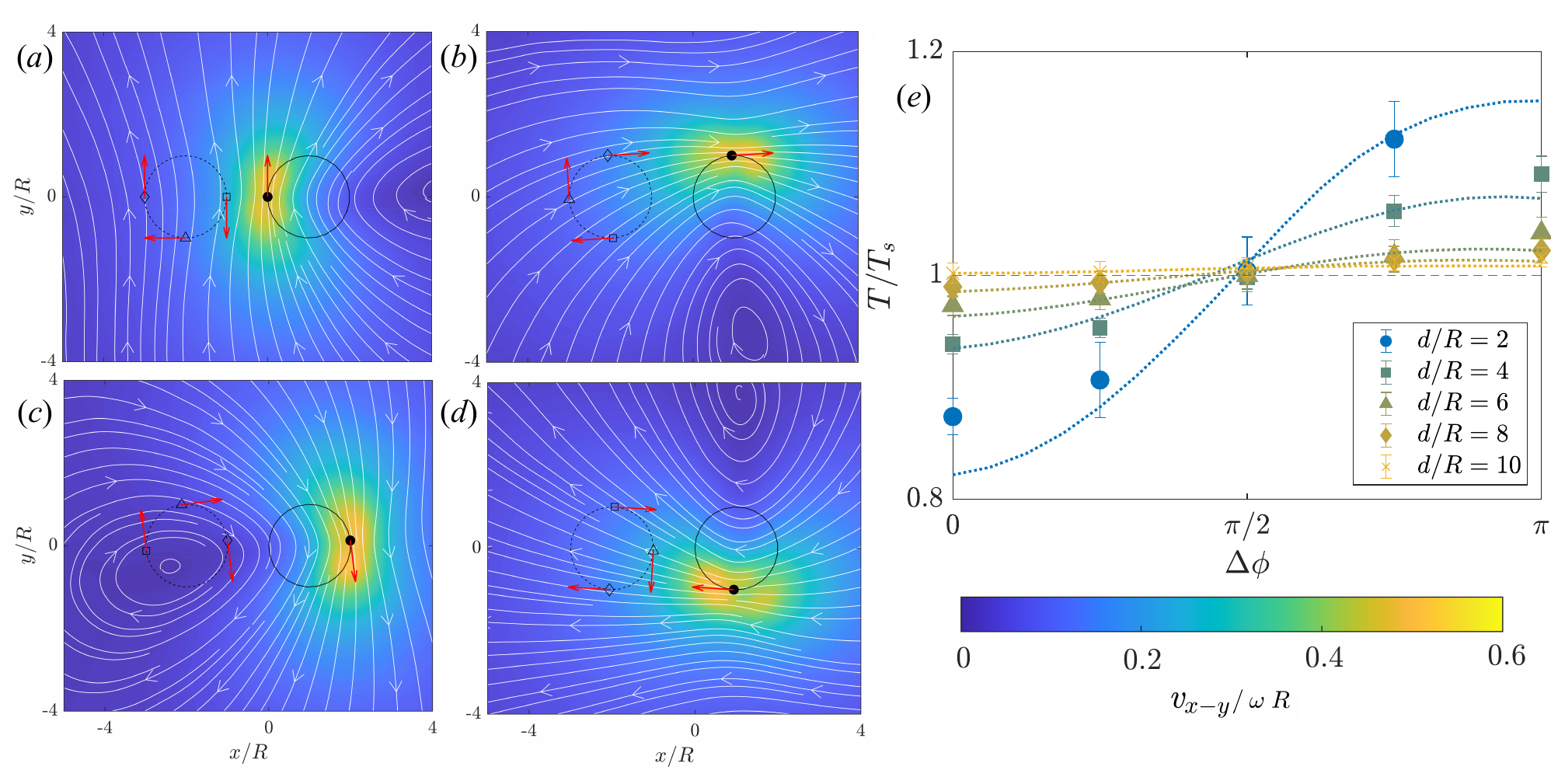}
\captionsetup{width=1\linewidth,justification=justified}
\caption{Hydrodynamic interactions between rotating helices. (a-d) The $x$-$y$ flow field of a single rotating helix with its axis positioned at $(R,0)$ over one rotation cycle obtained from PIV. White lines indicate streamlines, while the color represents the magnitude of the in-plane velocity $v_{x-y}$, normalized by the rotation speed $\omega R$ (see colorbar below panel (e)). The solid black line represents the circular trajectory of the real helix, with its cross-section in the plane marked by a black dot. When an imaginary helix is placed at $(-2R, 0)$, it also traces a circular path, shown as the dashed circle. The empty symbols mark the position of the imaginary helix when the phase difference between the real and imaginary helices is $\Delta\phi = 0$ (diamond, left), $\pi$/2 (triangle, bottom), and $\pi$ (square, right), respectively. Red arrows indicate the rotation velocity of the real and imaginary helices. (e) Torque on an imaginary helix in the flow field of a real helix, $\langle v_\theta^r\rangle/\omega R$, approximates the torque on one helix in the helical pair, $T/T_s$. Symbols are experimental measurements of $T/T_s$, the same as those in Fig.~\ref{fig:torque}. Lines represent $\langle v_{\theta}^r \rangle / \omega R$.}
\label{fig:bigtank}  
\end{figure}

\subsection{Hydrodynamic interaction} \label{HI_torque}
To illustrate the effect of hydrodynamic interactions between two rotating helices on the torque, we first show the PIV flow field of a \textit{single} rotating helix in the $x$-$y$ plane (Fig.~\ref{fig:bigtank}(a)). Next, we introduce an imaginary helix, placed at a distance $d = 3R$ to the left of the real helix, that experiences but does not disturb the flow of the real helix. This imaginary helix enables a `one-way' hydrodynamic coupling, where the imaginary helix experiences the flow and the resulting hydrodynamic force from the real helix, but does not influence the real helix. Consider three phase differences, $\Delta \phi = 0$, $\pi/2$, and $\pi$, represented by the three positions of the imaginary helix (three empty symbols along the dashed circle in Fig.~\ref{fig:bigtank}(a)). At $\Delta \phi = 0$ (empty diamond on the left), the flow of the real helix aligns with the rotation direction of the imaginary helix. As a result, the imaginary helix rotates at a lower speed relative to the local flow, experiencing reduced fluid drag and, consequently, smaller torque. In contrast, at $\Delta \phi = \pi$ (empty square on the right), the local flow opposes the rotation direction of the imaginary helix. The imaginary helix rotates at a higher speed relative to the local flow, resulting in increased drag and larger torque. Finally, at $\Delta \phi = \pi/2$ near $\Delta \phi_c$ (empty triangle at the bottom), the flow is nearly perpendicular to the rotation direction of the imaginary helix, causing minimal modification to the torque experienced by the imaginary helix. As the helices rotate, their relative positions evolve over time (Fig.~\ref{fig:bigtank}(a)-(d)). Nevertheless, throughout the rotation cycle, the relative orientation between the rotation of the imaginary helix and the local flow remains qualitatively similar, except when a flow vortex passes near the imaginary helix (Fig.~\ref{fig:bigtank}(c)). During this phase, however, the magnitude of the local flow drops significantly due to the presence of the vortex, minimizing the hydrodynamic interaction between the helices.

Quantitatively, we calculate the tangential velocity of the imaginary helix in the reference frame of the local flow of the single real helix at time $t$, $v_{\theta}^r(t) = \omega R - v_l(t)$, where $\omega R$ is the rotation velocity of the imaginary helix in the lab frame and $v_l(t)$ is the local flow velocity of the real helix projected onto the rotation direction of the imaginary helix. We then take a time average over one rotation period: $\langle v_{\theta}^r \rangle = (\omega/2\pi)\int_0^{2\pi/\omega} v_{\theta}^r(t)dt$. Due to the periodicity of the helices, this cycle average at a fixed height $z$ is equivalent to a spatial average over a pitch at a fixed time $t$, apart from minor end effects. 
As the torque on a helix is proportional to its rotation speed, the torque on the imaginary helix in the reference frame of the local flow of the real helix, relative to the torque of the real helix in the lab frame, is $\langle v_{\theta}^r \rangle / \omega R$, which serves as a good estimate of the normalized torque on a helix in the helical pair, $T/T_s$ (Fig.~\ref{fig:bigtank}(e)).

\subsection{Confinement}
\begin{figure}
  \centerline{\includegraphics[scale=0.4]{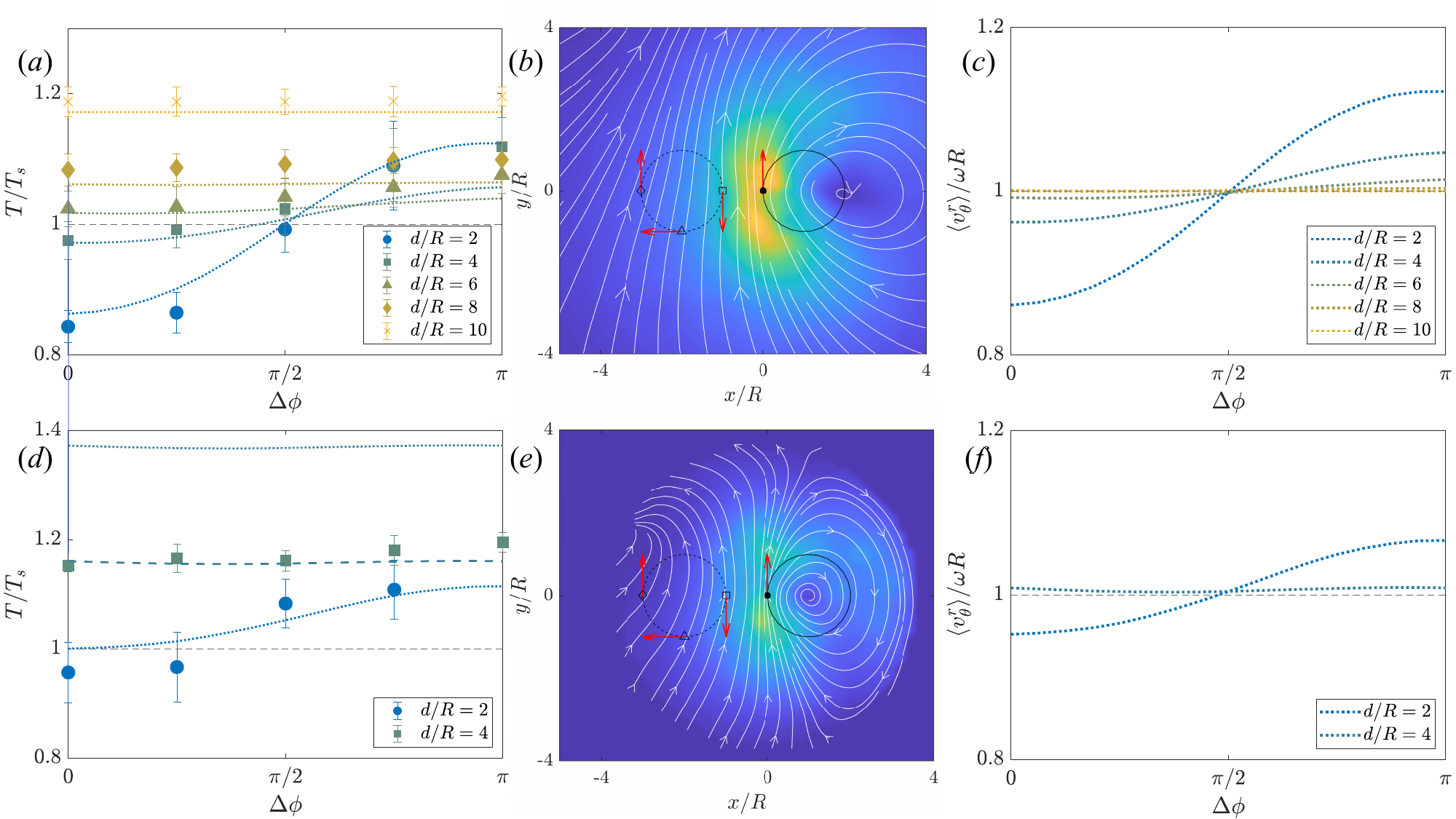}}
  \captionsetup{width=1\linewidth,justification=justified}
  \caption{Hydrodynamic interactions between rotating helices under confinement $R_0 = 7R$ (a-c) and $R_0 = 3.5R$ (d-f).  (a, d) Torque on one helix in the helical pair as a function of phase difference $\Delta \phi$ at different spacings $d$. Symbols denote experimental data, while dotted lines represent model predictions (Eq.~\eqref{eq:totaltorque}). The green dashed line in (d) is the model prediction using experimentally obtained $T_e/T_c$, instead of Eq.~\eqref{eq:cylinder}. (b, e) Flow field in the $x$-$y$ plane of a rotating helix positioned at $(R,0)$. The circle on the right indicates the helix’s trajectory, while the dashed circle on the left represents the trajectory of an imaginary helix. Three phase differences between the real helix and the imaginary helix are shown, consistent with those in Fig.~\ref{fig:bigtank}(a). (c, f) Tangential velocity of the imaginary helix in the reference frame of local flow, $\langle v_\theta^r \rangle$, normalized by the helix’s rotation speed in the lab frame, $\omega R$, as a function of $\Delta \phi$ at different $d$.}
\label{fig:confinement}
\end{figure}

Bacteria often inhabit confined or structured environments, such as soil pores, intestines, and biofilms, where their motility is significantly altered \citep{Vizsnyiczai2020}. To understand the pumping and mixing of rotating flagella in narrow microfluidic channels, it is also important to resolve the dynamics of rotating helices under confinement. However, a SBT formulation for rotating helices in cylindrically confined geometries remains unavailable. Instead, we experimentally investigate a pair of rotating helices under confinement using tubes of varying radii $R_0$ and extend our simple model of hydrodynamic interactions to predict the torque on confined helices.  

Figures~\ref{fig:confinement}(a) and (d) show the torque on one helix in the helical pair as a function of phase difference at varying spacings under two different confinement conditions. Similar to the weakly confined system, the torque increases with $\Delta \phi$ and this increasing trend weakens at large spacings. However, under confinement, the mean torque shifts to higher values as $d$ increases. Consequently, the critical phase difference $\Delta \phi_c$, at which the hydrodynamic interactions disappear, decreases with increasing $d$ and eventually vanishes at large $d$.  

Following the method in Sec.~\ref{HI_torque}, we estimate the torque on the helices from the flow field of a single helix located near the center of the confining tube (Figs.~\ref{fig:confinement}(b) and (e)). The cycle-averaged relative tangential velocity of a rotating imaginary helix in the flow field of the single helix $\langle v_\theta^r \rangle$ is computed at different $\Delta \phi$ and $d$ (Figs.~\ref{fig:confinement}(c) and (f)). While $\langle v_\theta^r \rangle/\omega R$ captures the trend of $T/T_s$, it fails to predict the upward shift in the mean torque with increasing $d$. As a result, the predicted $\Delta \phi_c$ remains close to $\pi/2$ regardless of confinement.

\begin{figure}
  \centerline{\includegraphics[scale=0.32]{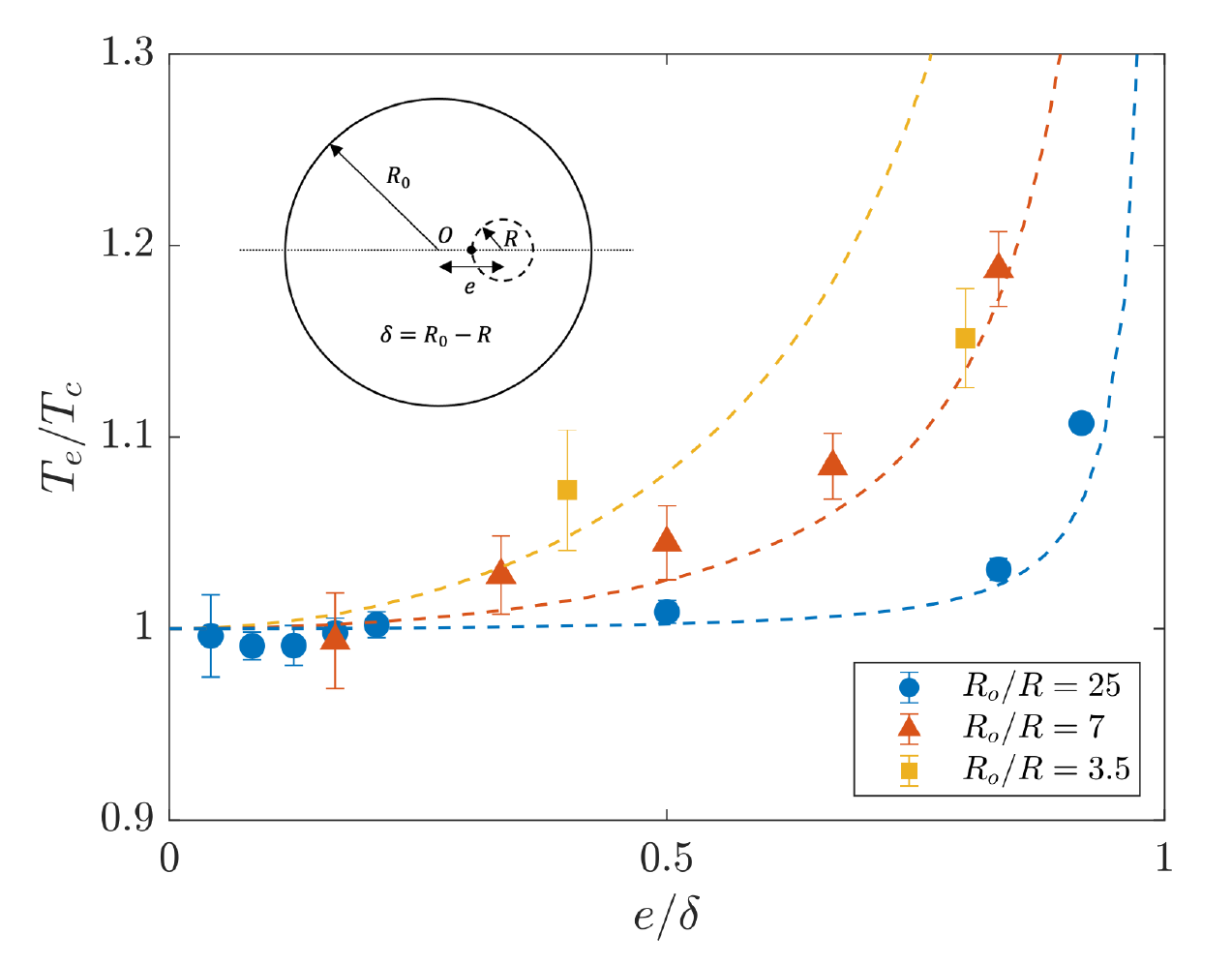}}
  \captionsetup{width=1\linewidth,justification=justified}
  \caption{Torque on a cylinder or helix $T_e$ in tubes of radius $R_0$ as a function of the eccentricity $e/\delta$. $T_e$ is normalized by the torque of a helix rotating concentrically in the tubes, $T_c$. Symbols are from experiments and lines are from Eq.~\eqref{eq:cylinder}. Inset: Schematic illustrating the position of the cylinder/helix within a tube.}
\label{fig:eccentricityplot}
\end{figure}

The offset of the measured mean torque arises because of the eccentric positioning of the helices and their proximity to the system boundary at large $d$ in confined systems. As the effect of the shape of an object on fluid drag is significantly subdued in Stokes flow, we hypothesize that the cycle-averaged torque on a rotating helix within a tube can be approximated by the torque on a rotating cylinder in the tube, where the radius of the cylinder is equal to the helical radius of the helix $R$. The ratio between the torque on an eccentric cylinder $T_e$ to that on a concentric cylinder $T_c$ is given as \citep{Jeffery1922,Yamagata1970}
\begin{equation} \label{eq:cylinder}
\frac{T_e}{T_c}=\frac{1}{G} \frac{P}{Q} \frac{1}{1+(e / \delta)^2 / 2} \quad \quad \text{with}
\end{equation}
\begin{equation}
P=1+2(e / \delta)^2 \frac{1+(\delta / 2 R)-(\delta / 2 R)^2}
{(1+\delta / R)^2}-\frac{1}{2} \frac{(e / \delta)^4(\delta / R)^2}{(1+\delta / R)^2}, \nonumber 
\end{equation}
\begin{equation}
Q=\left[1-\frac{(e / \delta)^2}{\left[1+\frac{1}{2}(\delta / R)(1-e / \delta)^2\right]^2}\right]^{\frac{1}{2}}, \quad \mathrm{and}\quad G=1-\frac{1}{2}(e / \delta)^2 \frac{\delta / R}{1+(\delta / 2 R)}. \nonumber
\end{equation}
Here, $\delta = R_0-R$ is the shortest distance between the center of the cylinder to the wall of the tube and $e/\delta = d/(2\delta)$ is the eccentricity (Fig.~\ref{fig:eccentricityplot} inset). To verify our hypothesis, we first identify the critical phase difference $\Delta \phi_c$ at each $d$ and $R_0$ from our calculation of $\langle v_\theta^r \rangle$ (Figs.~\ref{fig:confinement}(c) and (f)), where the hydrodynamic interactions between the two helices are nullified. We then obtain $T / T_s$ at $\Delta \phi_c$ from experiments via interpolation (Fig.~\ref{fig:confinement}(a) and (d)) and compare the results with the prediction of Eq.~\eqref{eq:cylinder} (Fig.~\ref{fig:eccentricityplot}). Experiments and the model exhibit good quantitative agreement in the weakly and moderately confined geometries, supporting our hypothesis. Not surprisingly, Eq.~\eqref{eq:cylinder} overestimates the torque by $\sim 20\%$ for the most confined geometry when the helix is too close to the wall at large $d$, where a rotating helix can no longer be accurately approximated by a rotating cylinder. 

The quantitative agreement shown in Fig.~\ref{fig:eccentricityplot} suggests that we can decompose the effect of the confining boundary on the torque of rotating helices into two parts. (\textit{i}) The boundary modifies the flow field of the helices, altering their mutual hydrodynamic interactions, as reflected in the trend of $\langle v_\theta^r\rangle/\omega R$. (\textit{ii}) The helices experience additional drag, $(T_e-T_c)/T_c$, from the hydrodynamic interactions with the boundary due to their eccentric positions within the confining tube. This decomposition of the confinement-induced hydrodynamic interactions into pair-induced and self-induced contributions is similar to the approach adopted in a recent study on hydrodynamic bound states of rotating cylinders under confinement \citep{Guo2024}. By summing the two contributions 
\begin{equation} \label{eq:totaltorque}
    \frac{T}{T_s} = \frac{\langle v_\theta^r\rangle}{\omega R} + \frac{T_e-T_c}{T_c},
\end{equation} 
we quantitatively predict the torque on a pair of rotating helices across different spacings, phase differences, and confinement levels (Figs.~\ref{fig:bigtank}(e), \ref{fig:confinement}(a) and (d)).  

What are the implications of our findings on bacteria swimming under confinement? First, experiments on \textit{E. coli} show that bacteria tend to swim near solid boundaries in channels of large radii but shift to swimming along the central axis of more confined channels when the confining radius $R_0$ is less than $\sim 7R$ \citep{Vizsnyiczai2020}. Our results suggest that this axial swimming behavior, characterized by low eccentricity, reduces the drag coefficient of the flagellar bundle $\xi$ in the torque-speed relation $T=\xi \omega$ (Figs.~\ref{fig:confinement}(a) and (d)). Thus, in confined environments, a bacterium is situated at a location, where its flagella rotate most rapidly under constant motor torque. Second, our findings reveal that the change of $\xi(d)$ at $\Delta \phi = 0$ becomes more pronounced in confined systems. This implies that the speed-up of flagellar rotation during bundle formation discussed above should be more evident under confinement---an intriguing prediction worth testing experimentally. Lastly, although our results support prior predictions that $\xi$ increases with the degree of confinement \citep{Vizsnyiczai2020}, our data do not clarify whether confinement enhances or reduces the thrust force of a flagellar bundle \citep{Liu2014}. For a translationally fixed rotating helix subjected to a constant torque, the resulting thrust force is $F = cT/\xi$, where $c$ is the off-diagonal term in the drag coefficient matrix that couples the thrust and rotation of the helix, $F = c\omega$ \citep{Kamdar2022_nature,Kamdar_2023_multipleflagella}. While our study shows that $\xi$ increases with confinement, $c$ also increases concurrently \citep{Vizsnyiczai2020}. Future work should aim to directly measure the thrust force of a helical pair under confinement, enabling an estimate of $c(R_0)$.

\begin{figure}
    \centerline{\includegraphics[scale=0.4]{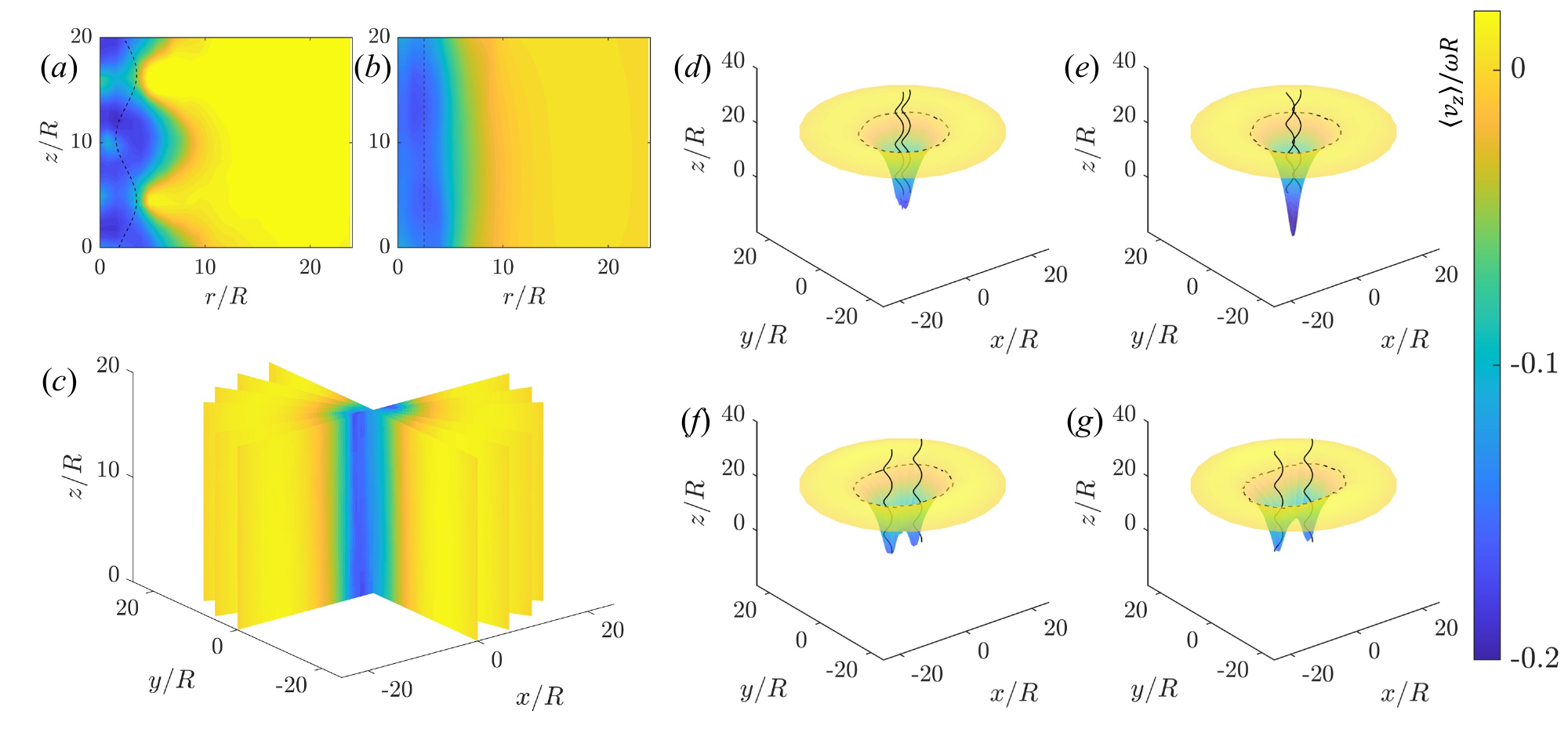}}
    \captionsetup{width=1\linewidth,justification=justified}
    \caption{Three-dimensional axial flow field from experiments. (a) Instantaneous $v_z(r,z)$ when the cross-sections of the helices at the $z=0$ plane are oriented at $\phi_1 = \phi_2 =1.28\pi$ and (b) time-averaged axial flow field $\langle v_z(r,z)\rangle$ over multiple rotation cycles for $\Delta\phi = 0$ and $d/R = 6$ in the axial plane ($\theta = 0$). (c) Stack of seven time-averaged axial flow fields at different $\theta$. 3D axial flow field with $d/R = 2$ and $\Delta\phi = 0$ (d); $d/R = 2$ and $\Delta\phi$ = $3\pi/4$ (e); $d/R = 10$ and $\Delta\phi = 0$ (f); and $d/R = 10$ and $\Delta\phi$ = $\pi$ (g). Dashed circles indicate $r_{c}(\theta)$ where $\langle v_z \rangle =0$.}
\label{fig:3dflowfield}
\end{figure}

\subsection{Axial flux and pump efficiency}

Finally, we investigate the fluid transport capacity of two rotating helices in the weakly confined system ($R_0=25R$) by imaging the 3D axial flow. While the radial flow $v_r$ in the axial plane quantitatively agrees with our SBT calculation and previous numerical predictions \citep{Kim2004_1, buchmann2018} (Movie 2), the axial flow $v_z$ exhibits qualitatively different behavior (Movie 1). Figures~\ref{fig:3dflowfield}(a) and (b) show an instantaneous and the time-averaged flow fields of the axial flow through the axes of the helices, which reveal a downward flow $(-z)$ near the helices and an upward flow $(+z)$ near the boundary. Unlike in an unbounded fluid, where $v_z$ decays to zero as $r\to\infty$, $v_z$ reaches zero at a finite radius $r_c$ and becomes positive near the wall. The confinement inherent in any experimental system ensures that the net flux across any cross-section of the system normal to the axes of the helices is zero. Due to the lack of the axisymmetry in the flow field, the locus $r_c(\theta)$ is non-circular and varies with the azimuthal angle $\theta$. We determine $r_c(\theta)$ for each experiment by interpolating the time-averaged flow field to identify the contour where $\langle v_z \rangle = 0$.

\begin{figure}
 \includegraphics[width=1\linewidth]{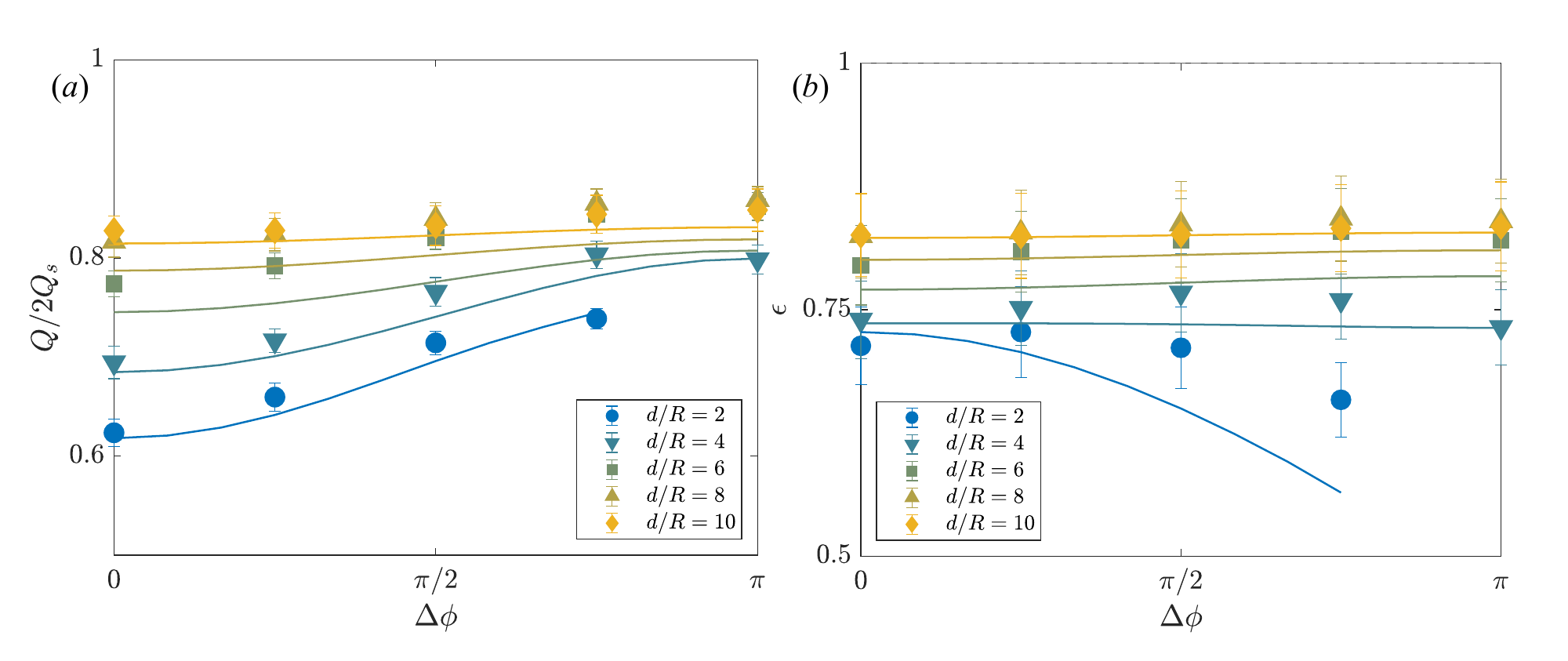}
 \captionsetup{width=1\linewidth,justification=justified}
 \caption{Fluid transport and pump efficiency. (a) Normalized axial flux of rotating helices as a function of phase difference $\Delta\phi$ for various spacings $d$. (b) Pump efficiency $\epsilon$ as a function of $\Delta\phi$ for different $d$. Symbols represent experimental data, while lines correspond to SBT calculations.}\label{fig:fluxcompandeff}
\end{figure}

To image the 3D flow field, we rotate the supporting rail and change the polar angle $\theta \in [0, \pi/2]$ (Figs.~\ref{fig:setup}(c) and \ref{fig:3dflowfield}(c)), leveraging the bilateral symmetry of the system. Representative time-averaged 3D axial flow fields, $\langle v_z(r,\theta) \rangle$, for helices of two phase differences and spacings are shown in Fig.~\ref{fig:3dflowfield}(d-g). We compute the axial flux $Q = \int_0^{2\pi} \int_0^{r_c(\theta)} \langle v_z(r, \theta)\rangle rdr d \theta$ from these flow fields (Fig.~\ref{fig:fluxcompandeff}(a)). Compared with the flux of two independently rotating helices $2Q_s$, helices pump less fluid when acting in a bundle, regardless their phase difference $\Delta \phi$ or spacing $d$. Similar to torque, $Q$ increases $\Delta \phi$ with the increasing trend diminishing at large $d$. For two helices at the smallest spacing in our experiments, $Q \approx 1.2Q_s$ when $\Delta \phi = 0$, indicating that the two closely spaced helices behave similarly to a single rotating helix. In an unbounded fluid, SBT predicts a diverging flux due to the slow decay of the axial flow following $\sim 1/r$ at large distances. However, the ratio of the flux of two helices to that of a single helix remains finite, as the pair effectively behaves like a single helix at large distances. Interestingly, despite the qualitative difference in the flow field, $Q/2Q_s$ from SBT provides a good match to experimental results (Fig.~\ref{fig:fluxcompandeff}(a)).

Although helices in a bundle pump less fluid, they also require less power to rotate. We quantify this trade-off by measuring the pump efficiency $\epsilon = [Q/(2T\omega)]/[Q_s/(T_s\omega)]$ \citep{Barrett2024}. Figure~\ref{fig:fluxcompandeff}(b) shows $\epsilon$ for a pair of helices at different $\Delta \phi$ and $d$. At small spacings, the pair are more efficient when rotating in phase. However, their efficiency always remains lower than that of two independently rotating helices. The SBT calculation provides a reasonable estimate of the experimental results.

A swimmer and a pump represent dual manifestations of the same low-$\Rey$ fluid phenomenon: a tethered swimmer functions as a pump, while an unanchored pump behaves as a swimmer \citep{Raz_2007_pumpswimming}. Specifically, the fluid flux generated by fixed rotating helices scales linearly with their swimming speed when untethered, while the power required for rotation as a pump is linearly correlated with the power needed for swimming. Thus, our results suggest that a bundle of helical filaments is less efficient for swimming than individual filaments rotating at the same speed.

\section{Conclusion}\label{notstyle}
By integrating experiments with hydrodynamic modeling, we present a study on the dynamics of rotating helices in low-Reynolds-number flow---a fluid phenomenon with fundamental implications for the function of bacterial flagellar bundles and practical relevance for microfluidics. In particular, we construct a scaled experimental model of a flagellar bundle of two helices and examine how helix spacing and phase difference influence torque, flow fields, and fluid transport. Building upon the flow characteristics of a single helix, we develop a simple model that quantitatively predicts the torque of a pair of rotating helices.  We further extend this model to the complex yet important case of rotating helices under confinement, enabling torque calculations that are otherwise inaccessible through existing analytical theories. Finally, we explore the fluid transport capabilities of rotating helices and assess their efficiency as low-Reynolds-number fluid pumps. Our findings provide key insights into the dynamics of bacterial flagellar bundles and offer valuable guidelines for designing microfluidic pumps.

\backsection[Supplementary data]{\label{SupMat}Supplementary movies are available at ...}

\backsection[Acknowledgements]{We thank Daniel Retic and Aidan Dosch for their help with the experimental setup, and Maria Tătulea-Codrean, Dipanjan Ghosh, Sookkyung Lim, and Eric Lauga for fruitful discussions.}

\backsection[Funding]{The research was supported by NSF CBET 2415405 and University of Minnesota MRSEC.}

\backsection[Declaration of interests]
{The authors report no conflict of interest.}

\appendix
\section{Slender-body theory}
\label{appendix:SBT}
Slender-body theory (SBT) provides a widely used mathematical framework for analyzing the dynamics of flagella and cilia in microorganism locomotion \citep{Lauga_2020_book}. We build on the SBT formulation developed by \citet{rodenborn2013}, which relates the force density along a rigid slender helix to its geometry, as well as its rotational and translational velocities in an unbounded fluid. Here, we extend their model to include a second, rotating helix that is parallel to the original and separated from it by a distance $d$. The choice of SBT is supported by the work of \citet{martindale2016}, which compared several computational methods for modeling helical filaments. The study shows that SBT is the optimal choice, offering a balance of accuracy and computational efficiency for the parameter regime relevant to our system with $\lambda/R = 12$ and $a/R = 0.1$. Here, $\lambda$ and $R$ are the pitch length and radius of the helix, respectively, and $a$ is the radius of the filaments.

We discretize each helix by a set of nodal points $\{\varphi_n\}$ along the helical angle $\varphi(s) = (2\pi/\lambda)s\cos\theta$, where $\theta = \text{atan}(\lambda/2\pi R)$ is the pitch angle and $s$ is the arc length. 
Each helix is approximated by $N$ straight segments of length $\Delta s = R\Delta\varphi\csc\theta$, where $\Delta\varphi$ is the mesh size in the helical phase. Following the parameters used by \citet{rodenborn2013}, we set $N = 54$, yielding a helix of finite length $L = 32R$ that matches the geometry of the helices in our experiments. 

The centerline of \emph{Helix 1} $(n=1,\dots,N)$ is 
\begin{equation}
\label{eq:1}
\mathbf{r}_1(\varphi_n) = R\left(\cos\varphi_n -\frac{d}{2R},-\sin\varphi_n,\varphi_n\cot\theta\right),
\end{equation}
whereas that of the \emph{Helix 2} $(n=N+1,\dots,2N)$ is
\begin{equation}
\label{eq:2}
\mathbf{r}_2(\varphi_n) = R\left(\cos\varphi_n + \frac{d}{2R},-\sin\varphi_n,\varphi_n\,\cot\theta\right).
\end{equation}
Using the formulation from Eq.~(18) of \citet{rodenborn2013}, extended to include to the second helix with $n=1,\dots,2N$, the velocity at node $n$ is 
\begin{equation}\label{eq:3}
\mathbf{u}_n = 
\frac{\bigl[\mathbf{I} - \hat{\mathbf{t}}_n\hat{\mathbf{t}}_n^{T} + \mathbf{K}_n\bigr]\cdot \mathbf{f}_n}{4\pi \eta} +
\frac{\Delta s}{8\pi \eta}
\sum_{\substack{m=1 \\ m\neq n}}^{2N}
\bigl(\frac{\mathbf{I}}{\lvert \mathbf{r}_{nm} \rvert} + \frac{\mathbf{r}_{nm}\mathbf{r}_{nm}^{T}}{\lvert \mathbf{r}_{nm} \rvert ^3}\bigr) \cdot \mathbf{f}_m,
\end{equation}
where $\mathbf{r}_n=\mathbf{r}_{k(n)}(\varphi_n)$ is the centerline position of node $n$, $\mathbf{r}_{nm} = \mathbf{r}_{k(n)}(\varphi_n) - \mathbf{r}_{k(m)}(\varphi_m)$ is the relative position between two segments $n$ and $m$ and $k(n) = 1$ or 2 indicates whether segment $n$ belongs to \textit{Helix 1} or \textit{Helix 2}. The first term denotes local contribution, where $\hat{\mathbf{t}}_n$ is the local tangent at node $n$, $\mathbf{f}_n$ is the unknown force density, and $\mathbf{K}_n=\ln \frac{\Delta s}{2\delta^{'}}\bigl(\mathbf{I}+ \hat{\mathbf{t}}_n\hat{\mathbf{t}}_n^{T}\bigr)$ arises from an asymptotic integral over a small neighborhood of node $n$ between the ``natural cutoff'' $\delta^{'} = a\sqrt e/2$ and $\Delta s/2$ \citep{Lighthill1976}. Here, $e$ is Euler's number. The second term accounts for the far-field contribution of all other segments $m\neq n$, including those on the other helix. Equation~\eqref{eq:3} can be expressed more compactly as a linear mapping: \begin{equation}\label{eq:5}
     \mathbf{u}^T = \mathbf{G}\mathbf{f}^T,
\end{equation}
where $\mathbf{u} = (\mathbf{u}_1,\mathbf{u}_2 \ldots \mathbf{u}_{2N})$, $\mathbf{f} = (\mathbf{f}_1,\mathbf{f}_2 \ldots \mathbf{f}_{2N})$, and  $\mathbf{G}$ is a $6N \times 6N$ matrix that captures both self- and mutual hydrodynamic interactions. The diagonal terms account for each segment's self-interaction via the local dipole correction, while the off-diagonal terms represent interactions between all segment pairs $(n, m)$, whether on the same helix or across different helices.

For a prescribed motion---such as rigid-body rotation about a fixed axis with angular velocity $\mathbf{\Omega}$, combined with a translation velocity $\mathbf{V}$---the nodal velocity $\mathbf{u}_n$ is determined by the no-slip boundary condition:
\begin{equation}\label{eq:4}
\mathbf{u}_n = \mathbf{\Omega} \times \bigl(\mathbf{r}_n-\mathbf{r}_{\text{axis}(n)}\bigr)+ \mathbf{V}\,
\end{equation}
where $\mathbf{r}_{\text{axis}(n)}$ is the center of rotation of the helix to which segment $n$ belongs. By equating Eq.~\eqref{eq:5} and \eqref{eq:4}, we solve for the force densities $\{\mathbf{f}_n\}_{n=1...2N}$. The axial force $F_z$ and torque $T$ on each helix are obtained by summing over the corresponding subset of nodes.  


Finally, the flow field $\mathbf{v(x)}$ is evaluated from $\{\mathbf{f}_n\}$ by \citep{Kim2004_1}
\begin{equation}
v_i(\mathbf{x})=\sum_{n=1}^{2 N}\left[S_{i j}\left(\mathbf{x}, \mathbf{r}_n\right) \frac{f_{n j}}{8 \pi \eta}-D_{i j}\left(\mathbf{x}, \mathbf{r}_n\right) \frac{a^2 f_{n, \perp j}}{16 \pi \eta}\right] \Delta s_n
\end{equation}
where $i$ and $j$ run over $x$, $y$ and $z$. The summation runs over the arc length of both helices. $\mathbf{f}_n$ is the force per unit length exerted on the fluid by the $n$th segments and $\mathbf{f}_{n \perp}$ is the transverse part of $\mathbf{f}_n$. The Stokeslet tensor 
\begin{equation}
S_{i j}\left(\mathbf{x}, \mathbf{r}_n\right)=\frac{\delta_{i j}}{\left|\mathbf{x}-\mathbf{r}_n\right|}+\frac{\left(x_i-r_{n, i}\right)\left(x_j-r_{n, j}\right)}{\left|\mathbf{x}-\mathbf{r}_n\right|^3},
\end{equation}
and the doublet tensor  
\begin{equation}
D_{i j}\left(\mathbf{x}, \mathbf{r}_n\right)=-\frac{\delta_{i j}}{\left|\mathbf{x}-\mathbf{r}_n\right|^3}+3 \frac{\left(x_i-r_{n, i}\right)\left(x_j-r_{n, j}\right)}{\left|\mathbf{x}-\mathbf{r}_n\right|^5}.
\end{equation}






\bibliographystyle{jfm}
\bibliography{jfm}

\end{document}